\documentclass[a4paper,12pt]{article}
\usepackage{amsmath}
\usepackage{amsfonts}
\usepackage[dvips,bookmarksnumbered=true,breaklinks=true]{hyperref}
\usepackage{a4}
\usepackage{parskip}
\newcommand{\inv}[1]{\frac{1}{#1}}
\newcommand{\starco}[2]{\left[#1\stackrel{\star}{,}#2\right]}
\newcommand{\staraco}[2]{\left\{#1\stackrel{\star}{,}#2\right\}}

\newcommand{\var}[2]{\frac{\d #1}{\d #2}}

\newcommand{\intx}{\int d^4x}

\newcommand{\Gam}{\Gamma^{(0)}}
\newcommand{\Act}{\Gamma}
\newcommand{\ri}{{\rm i}}
\newcommand{\rig}{{\rm i}g}

\renewcommand{\d}{\delta}

\renewcommand{\th}{\theta}

\newcommand{\p}{\tilde{p}}

\newcommand{\bc}{\bar{c}}
\newcommand{\tc}{\widetilde{c}}

\newcommand{\St}{\Act_{\rm tot}}

\newcommand{\x}{\tilde{x}}

\newcommand{\uim}{UV/IR mixing }
\newcommand{\nc}{non-commutative }
\title{Non-Commutative \texorpdfstring{$U(1)$}{U(1)} Gauge Theory on \texorpdfstring{$\mathbb{R}^4_{\Theta}$}{R4} with Oscillator Term and BRST Symmetry}
\author{Daniel N. Blaschke\footnotemark[1]~, Harald Grosse\footnotemark[2]~ and Manfred Schweda\footnotemark[1]}
\date{January 28, 2008}
\begin{document}
\maketitle
\begin{center}
\renewcommand{\thefootnote}{\fnsymbol{footnote}}
\vspace{-0.3cm}\footnotemark[1]Institute for Theoretical Physics, Vienna University of Technology\\Wiedner Hauptstrasse 8-10, A-1040 Vienna (Austria)\\[0.3cm]
\footnotemark[2]Faculty of Physics, University of Vienna\\Boltzmanngasse 5, A-1090 Vienna (Austria)\\[0.5cm]
\ttfamily{E-mail: blaschke@hep.itp.tuwien.ac.at, harald.grosse@univie.ac.at, mschweda@tph.tuwien.ac.at}
\vspace{0.5cm}
\end{center}
\begin{abstract}
Inspired by the renormalizability of the \nc $\phi^4$ model with added oscillator term, we formulate a \nc gauge theory, where the oscillator enters as a gauge fixing term in a BRST invariant manner. All propagators turn out to be essentially given by the Mehler kernel and the bilinear part of the action is invariant under the Langmann-Szabo duality. The model is a promising candidate for a renormalizable \nc $U(1)$ gauge theory.

\end{abstract}
\newpage
\section{Introduction}
Despite the successes of renormalized perturbation expansion of local quantum field theories, the convergence problem as well as the renormalon, respectively Landau ghost problem, still remain. Furthermore gravity is not included within the
Standard model of particle physics, which is formulated as a non-Abelian gauge model.

The idea to include certain gravity effects led to the formulation of \nc quantum field theory, were space-time is deformed. Although quantum group, Lie-algebra and canonical deformations are well-established, properties of quantum field theories defined over these spaces are well-studied only for the latter case.
These expansions suffer from the UV/IR mixing: Non-planar graphs show divergences for exceptional momenta in the infrared, which spoils the renormalizability of the models (cf.~\cite{Douglas:2001,Szabo:2001} and references therein for a review).

Through a careful study of the Euclidean scalar $\phi^4$ model on the deformed $\mathbb{R}^4_{\Theta}$ space, Wulkenhaar and one of the present authors (H.~G.) were able to solve this problem~\cite{Grosse:2003, Grosse:2004b}. The study of the renormalization group flow led to the identification of four relevant/marginal operators, one more than usual, an oscillator- like term was added. Two further proofs have been worked out in~\cite{Rivasseau:2005a,Rivasseau:2005b}.
The renormalization group flow for the coupling constant turned out to be bounded, which was shown by a first order calculation in \cite{Grosse:2004a}, extended to three loops in~\cite{Rivasseau:2006a} and to all orders recently by the Paris group~\cite{Rivasseau:2006b}. This might lead to a constructive
procedure for a \nc scalar $\phi^4$ theory. A recent review on this matter can be found in~\cite{Rivasseau:2007} and references therein.

\section{Towards a renormalizable \nc gauge theory}
Inspired by the renormalizability of \nc $\phi^4$ theory in $\mathbb{R}^4_{\Theta}$ with an oscillator term, we try to construct a renormalizable \nc $U(1)$ gauge theory, for simplicity also in Euclidean space. In $\phi^4$ theory, the propagator was modified by the oscillator term in such a way, that it became the Mehler kernel. Here, we try to do the same thing: Since an oscillator term $\intx \Omega^2\x^2A_\mu A_\mu$ is not gauge invariant, there are more or less two possible ways to construct the model: either one adds further terms in order to make the action gauge invariant (cf.~\cite{Wulkenhaar:2007,Grosse:2007}) or one views the oscillator term as part of the gauge fixing part of the action. Here, we take the latter approach and note, that the oscillator term has the form of a mass term with non-constant ``mass'' $m^2=\Omega^2\x^2$. However, we will add a further term to the gauge fixing action in order to simplify the gauge field propagator. Our starting point is hence the following action:
\begin{align}\label{action}
S&=\intx\left[\inv{4}F_{\mu\nu}F_{\mu\nu}-\inv{2}A_\mu\left(\partial_\mu\partial_\nu-\Omega^2\x^2\d_{\mu\nu}\right)A_\nu\right],
\end{align}
with
\begin{align}
F_{\mu\nu}&=\partial_\mu A_\nu-\partial_\nu A_\mu-\ri g\starco{A_\mu}{A_\nu},\nonumber\\
\x_\mu&=\left(\th^{-1}\right)_{\mu\nu}x_\nu,\nonumber\\
\ri\th_{\mu\nu}&=\starco{x_\mu}{x_\nu}.
\end{align}
$\th_{\mu\nu}$ is assumed to be a constant skew-symmetric matrix. Some remarks concerning this action are in order. Note that the Weyl-Moyal $\star$-product has the following properties:
\begin{align}\label{starprop}
\intx A_\mu(x) \star A_\nu(x)\star A_\rho ( x )&=\intx A_\rho ( x )\star A_\mu(x) \star A_\nu(x),\nonumber\\
\intx A_\mu(x) \star A_\nu(x)&=\intx A_\mu(x) A_\nu(x),\nonumber\\
\staraco{\x_\mu}{A_\nu(x)}&=2\x_\mu A_\nu(x).
\end{align}
Due to the last property, one may write for the oscillator term
\begin{align}\label{identity1}
\inv{4}\staraco{\x_\nu}{A_\mu}\star \staraco{\x_\nu}{A_\mu}=\left(\x_\nu A_\mu\right)\star\left(\x_\nu A_\mu\right)
\end{align}
and the remaining star is removed by the integral over space according to the second property of the star product. Hence, there are only ordinary products left in the oscillator term and $A$ and $\x$ may be rearranged to the form written in (\ref{action}).

From the bilinear part of the action (\ref{action}) we easily arrive at the equations of motion for the free fields:
\begin{align}
\var{S_\text{bi}}{A_\mu}&=\left(-\Delta_4+\Omega^2\x^2\right)A_\mu\,.
\end{align}
Notice that the terms $\partial_\mu\partial_\nu A_\nu$ have cancelled due to gauge fixing. The inverse of the operator $\left(\Delta_4-\Omega^2\x^2\right)$ gives the Mehler kernel which will become the propagator of the gauge field. Now we need to find the ghost sector for the action. In order to do this, we need to rewrite the gauge fixing in terms of some multiplier field and add ghosts.

Since our ``mass'' $\Omega^2\x^2$ is $x$-dependent, we cannot simply employ the gauge fixing and ghost sector of Curci and Ferrari~\cite{Ferrari:1976} (see also~\cite{Periwal:1995}). Instead, we suggest the following gauge fixed action in the classical limit:
\begin{align}\label{new-action}
\Gam&=\Act_{\text{inv}}+\Act_{\text{m}}+\Act_{\text{gf}}\,,\nonumber\\
\Act_{\text{inv}}&=\inv{4}\intx\, F_{\mu\nu}\star F_{\mu\nu}\,,\nonumber\\
\Act_{\text{m}}&=\frac{\Omega^2}{4}\intx\left(\inv{2}\staraco{\x_\mu}{A_\nu}\star\staraco{\x_\mu}{A_\nu}+\staraco{\x_\mu}{\bc}\star\staraco{\x_\mu}{c} \right)=\nonumber\\
&=\frac{\Omega^2}{8}\intx\left(\x_\mu\star\mathcal{C}_\mu\right)\,,\nonumber\\
\Act_{\text{gf}}&=\intx\Bigg[B\star\partial_\mu A_\mu-\inv{2}B\star B-\bc\star\partial_\mu sA_\mu-\frac{\Omega^2}{8}\,\tc_\mu\star s\,\mathcal{C}_\mu\Bigg]
\end{align}
with
\begin{align}
\mathcal{C}_\mu&=\Big(\staraco{\staraco{\x_\mu}{A_\nu}}{A_\nu}+\starco{\staraco{\x_\mu}{\bc}}{c}+\starco{\bc}{\staraco{\x_\mu}{c}}\Big)\,.
\end{align}
This action is invariant under the BRST transformations given by
\begin{align}\label{BRST}
&sA_\mu=D_\mu c=\partial_\mu c-\rig\starco{A_\mu}{c}, && s\bc=B,\nonumber\\
&sc=\rig{c}\star{c}, && sB=0,\nonumber\\
&s\tc_\mu=\x_\mu, && s^2\varphi=0\ \forall\ \varphi\in\left\{A_\mu,B,c,\bc,\tc_\mu\right\}.
\end{align}
$B$ is the multiplier field implementing the gauge fixing, which for $\tc_\mu\to0$ reduces to the usual covariant Feynman gauge $\partial_\mu A_\mu-B=0$. $\Omega$ is a constant parameter and $c$, $\bc$ are the ghost/antighost, respectively. The ``mass'' term for the ghosts (cf. second term in $\Act_{\text{m}}$) has been introduced in order to have a Mehler kernel also for the ghost propagator. The field $\tc_\mu$ is a multiplier field with mass dimension $1$ and ghost number $-1$, which imposes a constraint, namely on-shell BRST invariance of $\mathcal{C}_\mu$. In fact, because of $s\x_\mu=0$, this constraint also implies on-shell BRST invariance of the mass terms $\Act_{\text{m}}$. Furthermore, $s^2\,\mathcal{C}_\mu=0$ vanishes identically, i.e. off-shell. Using the properties of the star product (\ref{starprop}) and (\ref{identity1}), one may rewrite $\Act_{\text{m}}$ also in the form
\begin{align}
\Act_{\text{m}}&=\intx\,\Omega^2\x^2\left(\inv{2}A^2+\bc c\right)\,,
\end{align}
which is the most convenient for calculating the propagators.

A further comment we want to make, is that the classical action may be reexpressed by the formula
\begin{align}
\Gam&=\intx\left[\inv{4}F_{\mu\nu}\star F_{\mu\nu}+s\left(\bc\star\partial_\mu A_\mu\right)-\inv{2}B\star B+\frac{\Omega^2}{8}\,s\left(\tc_\mu\star\mathcal{C}_\mu\right)\right],
\end{align}
showing the unphysical character of the $s$-variations.

The bilinear parts of the action lead to the following improved propagators:
\begin{subequations}\label{propagators}
\begin{align}\label{propa}
G^{A}_{\mu\nu}(x-y)&=\left(-\Delta_4+\Omega^2\x^2\right)^{-1}\d_{\mu\nu}\d^4(x-y),\nonumber\\
G^{\bc c}(x-y)&=\left(-\Delta_4+\Omega^2\x^2\right)^{-1}\d^4(x-y),
\end{align}
\begin{align}\label{propb}
G^{BA}_\mu(x-y)&=\left(-\Delta_4+\Omega^2\x^2\right)^{-1}\partial_\mu\d^4(x-y),\nonumber\\
G^{B}(x-y)&=\left[\partial_\mu\left(-\Delta_4+\Omega^2\x^2\right)^{-1}\partial_\mu-1\right]\d^4(x-y).
\end{align}
\end{subequations}
Both the gauge field and the ghost propagators are essentially the Mehler kernel, so we may expect improved IR behaviour of the Feynman graphs. Since there are no vertices involving the $B$ field and since the additional multiplier $\tc_\mu$ has no propagator, neither field will play a role in loop corrections.

When adding external sources for the non-linear BRST transformations $sA_\mu$ and $sc$,
\begin{align}
\Act_{\text{ext}}=\intx\left[\rho_\mu\star sA_\mu+\sigma\star sc\right],
\end{align}
we arrive at the Slavnov-Taylor identity
\begin{align}
\mathcal{S}\left(\St\right)&=\intx\left(\var{\St}{\rho_\mu}\star\var{\St}{A_\mu}+\var{\St}{\sigma}\star\var{\St}{c}+B\star\var{\St}{\bc}+\x_\mu\star\var{\St}{\tc_\mu}\right)=0,
\end{align}
with
\[
\St=\Gam+\Act_{\text{ext}}.
\]
Finally, we also note the following observation: Using the equation of motion
\begin{align}
\var{\Gam}{B}&=\partial_\mu A_\mu-B+\frac{\Omega^2}{8}\Big(\starco{\staraco{\x_\mu}{c}}{\tc_\mu}-\staraco{\x_\mu}{\starco{\tc_\mu}{c}}\Big)=0
\end{align}
one may eliminate the $B$-field\footnote{This is equivalent to integrating out the $B$-field in the path integral.} from the action (\ref{new-action}). In that form, it becomes obvious, that the bilinear parts of the gauge fixed action are invariant under a Langmann-Szabo duality~\cite{Szabo:2002}. As usual after eliminating the $B$-field, the BRST transformation of $\bc$ is nilpotent only on-shell:
\begin{align}
s^2\bc=\var{\Act}{\bc}\,.
\end{align}

\section{Outlook}
In a first step (work in progress) we want to analyze the one-loop calculation of the vacuum polarization for the U(1)-photon with the presented concepts, in order to demonstrate that one is able to cure the \uim problem. With the improved Mehler propagators it is  expected that the dangerous UV/IR contributions
\begin{align}
\Pi^{\text{UV/IR}}_{\mu\nu}\propto\frac{\p_\mu\p_\nu}{(\p^2)^2}\,,\qquad\p_\mu=\th_{\mu\nu}\p_\nu\,,
\end{align}
which are gauge fixing independent, cancel. In a further step more general considerations (renormalization to all orders, RG-flow, etc.) are planned.

\section*{Acknowledgments}
D.~N. Blaschke is a recipient of a DOC-fellowship of the Austrian Academy of Sciences at the Institute for Theoretical Physics at Vienna University of Technology.\\
The work of M. Schweda was supported by "Fonds zur F\"orderung der Wissenschaftlichen Forschung" (FWF) under contract P19513-N16.


\end{document}